\documentclass{aa}
\usepackage{graphicx}
\usepackage{natbib}
\usepackage{epsfig}
\usepackage{txfonts}
\usepackage[figuresright]{rotating}
\usepackage{lscape}
\usepackage{hyperref}

\bibpunct{(}{)}{;}{a}{}{,}

\newcommand{\corot}{\textsl{CoRoT}}  

\def\kms{\,km\,s$^{-1}$}             
\def\vsini{$v$\,sin\,$i$}             
\def\m2s2{\hbox{\,m$^{2}$\,s$^{-2}$}} 
\def\kms{\hbox{\,km\,s$^{-1}$}}       
\def\gcm3{\hbox{\,g\,cm$^{-3}$}}      
\def\vsini{\hbox{$v$\,sin\,$i$}}      

\def\degr{\hbox{$^\circ$}}

\begin{document}

 \title{From CoRoT\,102899501 to the Sun \thanks{Based on  
 observations obtained with \corot, a space project operated by the
 French Space Agency, CNES, with participation of the Science
 Programme of ESA, ESTEC/RSSD, Austria, Belgium, Brazil, Germany and Spain.} 
 \fnmsep \thanks{Based on observations made with the Anglo-Australian Telescope; the 2.1-m Otto Struve telescope at McDonald Observatory, Texas, USA; the Nordic Optical Telescope, operated on the island of La Palma jointly by Denmark, Finland, Iceland, Norway, and Sweden, in the Spanish Observatorio del Roque de los Muchachos of the Instituto de Astrofisica de Canarias, in time allocated by the NOT ``Fast-Track'' Service Programme, OPTICON, and the Spanish Time Allocation Committee (CAT).} \fnmsep \thanks{The research leading to these results has received funding from the European Community's Seventh Framework Programme (FP7/2007-2013) under grant agreement
number RG226604 (OPTICON).}
}

\subtitle{A time evolution model of chromospheric activity on the main sequence}

   \author{
      P.~Gondoin \inst{1}, 
      D.~Gandolfi \inst{1,2}, 
      M.~Fridlund \inst{1}, 
      A.~Frasca \inst {3},
      E.\,W.~Guenther \inst{2}, 
      A.~Hatzes \inst{2},
      H.\,J.~Deeg \inst{4,5},
      H.~Parviainen \inst{4,5},
      P.~Eigm\"uller\inst{1,6},
\and  M.~Deleuil\inst{7}
          }

   \institute{European Space Agency, ESTEC - Postbus 299,
              2200 AG Noordwijk, The Netherlands
        \and Th\"uringer Landessternwarte Tautenburg,
             Sternwarte 5, 07778 Tautenburg, Germany
        \and INAF-Osservatorio Astrofisico di Catania,
             Via S.Sofia 78, 95123 Catania Italy
        \and Instituto de Astrof{\'i}sica de Canarias, 
             E-38205 La Laguna, Tenerife, Spain
        \and Departamento de Astrof\'isica, Universidad de La Laguna, 
             38206 La Laguna, Tenerife, Spain
        \and Institute of Planetary Research, German Aerospace Center, 
             Rutherfordstrasse 2, 12489 Berlin, Germany
        \and Laboratoire d'Astrophysique de Marseille, 
             38 rue Frédéric Joliot-Curie, 13388, Marseille Cedex 13, France 
     }

   \date{Received 23 February 2012 / Accepted 21 September 2012}

 
  \abstract {} 
{The present study reports measurements of the rotation period of a young solar analogue, estimates of its surface coverage by photospheric starspots and of its chromospheric activity level, and derivations of its evolutionary status. Detailed observations of many young solar-type stars, such as the one reported in the present paper, provide insight into rotation and magnetic properties that may have prevailed on the Sun in its early evolution.}
{Using a model based on the rotational modulation of the visibility of active regions, we analysed the high-accuracy \corot\ lightcurve of the active star CoRoT\,102899501. Spectroscopic follow-up observations  were used to derive its fundamental parameters. We compared the chromospheric activity level of Corot \,102899501 with the $R_{\rm HK}^{'}$ index distribution vs age established on a large sample of solar-type dwarfs in open clusters. We also compared the chromospheric activity level of this young star with a model of chrosmospheric activity evolution established by combining relationships between the $R_{\rm HK}^{'}$ index and the Rossby number with a recent model of stellar rotation evolution on the main sequence.}
{We measure the spot coverage of the stellar surface as a function of time, and find evidence for a tentative increase from 5-14\% at the beginning of the observing run to 13-29\% 35 days later. A high level of magnetic activity on Corot \,102899501 is corroborated by a strong emission in the Balmer and  \ion{Ca}{ii} H\,\&\,K lines ($log R_{\rm HK}^{'} \sim -4$). The starspots used as tracers of the star rotation constrain the rotation period to $1.625\pm0.002$~days and do not show evidence for differential rotation.\\
The effective temperature ($T_\mathrm{eff}=5180\pm80$\,K), surface gravity (log\,$g=4.35\pm0.1$), and metallicity ($[\rm{M/H}]=0.05\pm0.07$\,dex) indicate that the object is located near the evolutionary track of a 1.09$\pm$0.12~M$_\odot$ pre-main sequence star at an age of 23$\pm$10\,Myrs. This value is consistent with the ``gyro-age'' of about 8-25\,Myrs, inferred using a parameterization of the stellar rotation period as a function of colour index and time established for the $I$-sequence of stars in stellar clusters.} 
{We conclude that the high magnetic activity level and fast rotation of CoRoT\,102899501 are manifestations of its stellar youth consistent with its estimated evolutionary status and with the detection of a strong Li\,{\sc i}\,$\lambda$6707.8\,\AA\ absorption line in its spectrum. We argue that a magnetic activity level comparable to that observed on CoRot\,102899501 could have been present on the Sun at the time of planet formation.}


\titlerunning{CoRoT observation of a young Sun-like star}
\authorrunning{P. Gondoin et al.}
   \maketitle

\begin{table}[!t]
  \caption{CoRoT, 2MASS, and USNO-A2 identifiers of the target star. Equatorial coordinates, 
           optical, and near-infrared photometry are from the \emph{ExoDat} catalogue 
           (Deleuil et al. 2009) and \emph{2MASS} catalogue (Cutri et al. 2003).}
  \centering
  \begin{tabular}{lll}       

  \multicolumn{1}{l}{Main identifiers}     \\
  \hline
  \hline
  \noalign{\smallskip}                
   \corot~ID              & 102899501        \\
   \emph{2MASS}~ID        & 06483081-0234206  \\
   \emph{USNO-A2}~ID      & 0825-03232995     \\
  \noalign{\smallskip}                
  \hline
  \noalign{\medskip}
  \noalign{\smallskip}                

  \multicolumn{2}{l}{Coordinates}     \\
  \hline                                  
  \hline                                  
  \noalign{\smallskip}                
  R.A. (J2000)    & $~~06^h\,48^m\,30\fs81$         \\
  Dec (J2000)     & $-02\degr\,34\arcmin\,20\farcs53$  \\
  \noalign{\smallskip}                
  \hline
  \noalign{\medskip}
  \noalign{\smallskip}                

  \multicolumn{3}{l}{Magnitudes} \\
  \hline
  \hline
  \noalign{\smallskip}                
  \centering
  Filter & Mag & Error \\
  \noalign{\smallskip}                
  \hline
  \noalign{\smallskip}                
  $B$   & 13.727 & 0.047 \\
  $V$   & 12.846 & 0.056 \\
  $r^\prime$  & 12.512 & 0.056 \\
  $i^\prime$  & 11.934 & 0.065 \\
  $J$   & 11.095 & 0.026 \\
  $H$   & 10.570 & 0.022 \\
  $Ks$  & 10.461 & 0.021 \\
  \noalign{\smallskip}                
  \hline
  \end{tabular}
\label{Table_Names_Coord}      
\end{table}

\section{Introduction}

Cool stars generate magnetic fields through dynamo processes in their interiors. These fields reach the stellar photospheres, where they produce cool spots. Magnetic fields also control outer stellar atmospheres. They heat stellar coronae and produce flares whose by-products, such as shock waves and high-energy particles, interact with the atmospheres of planets. One major topic in studying stellar activity is to explain how these magnetic phenomena seen on the Sun and stars depend on stellar parameters and their evolution. 

Recent space-borne photometric missions such as \corot\ and $\it{Kepler}$ provide precision photometry for a large number of stars with different stellar properties and ages, making these lightcurves a powerful tool for understanding stellar magnetic activity. Although detailed analysis of only a small fraction of these lightcurves has been published, it has provided new information on rotation and differential rotation \citep{Lanza2009, Frohlich2009} as well as on the properties of spots \citep{Silva2010}, such as location, areal coverage, and lifetime \citep{Mosser2009}, for stars with different activity levels.

Part of the interest in magnetic phenomena comes from their possible impact on planet formation during the early phase of stellar evolution \citep{Gudel2007}. One question in particular concerns the level of magnetic activity on the Sun in its infancy, when planets and their atmospheres formed. Detailed observations of many young solar-type stars, such as the one reported in the present paper, will provide insight into the rotation and magnetic properties that may have prevailed on the Sun at the beginning of the solar system history \citep{Gaidos2000}.

In this study, we report on the analysis of the high-accuracy light curve of the active star CoRoT\,102899501 observed with the \corot\ satellite during its initial run in the exoplanet field IRa01 (Sect.~\ref{Sec:CoRoT Observations}). Its light curve exhibits spot-induced variability with a large amplitude and a short period that are indicative of high magnetic activity level coupled to rapid stellar rotation. These indicators are signs of stellar youth \citep{Simon1985,Gudel1997,Soderblom2001}, since rotation and magnetic activity on single late-type dwarfs decrease with stellar evolution.

The evolutionary status of CoRoT\,102899501 was derived from spectroscopic observations performed at the Anglo-Australian Observatory, McDonald Observatory, and Nordic Optical Telescope (Sect.~\ref{Sec:Spec FU}). The light curve analysis uses a model \citep{Lanza2006} based on the rotational modulation of the visibility of active regions (Sects.~\ref{Sec:Stellar variability model} and \ref{Sec:Lightcurve analysis}). Chromospheric activity levels and lithium abundance were assessed using a spectral subtraction technique (Sects.~\ref{Sec:Chrom} and \ref{Sec:Lithium}). Results are discussed in Sect.~\ref{Sec:Discussion}.

\section{Observations}

\subsection{CoRoT photometry}
\label{Sec:CoRoT Observations}

CoRoT\,102899501 was photometrically observed with the space telescope \corot\ (Baglin et al. 2006; Auvergne et al. 2009) during the initial run IRa01, from 6 February to 2 April 2007. The lightcurve is continuous over 54 days with a sampling time of 512 seconds along the entire observation. The passband of the photometric data used in the present study ranges from 350 to 1000 nm. Identifiers of the target are reported in Table~\ref{Table_Names_Coord}, along with its equatorial coordinates, optical and near-infrared magnitudes, as retrieved from the \emph{ExoDat} database \citep{Deleuil2009} and \emph{2MASS} catalogue \citep{Cutri2003}.

The pipeline reductions of the \corot\ lightcurve followed the scheme outlined by Barge et al. (2008). To detect and eliminate remaining outliers, we subtracted a moving-median filtered version of the reference lightcurve and flagged the points at distances greater than three times the dispersion of the residuals. These points were replaced by the median of previous and subsequent non-flagged values. Following the approach of \citet{Lanza2009}, we computed a filtered version of the lightcurve by means of a sliding median boxcar filter with a boxcar extension approximately equal to one orbital period of the satellite, i.e. 6184 seconds \citep[cf.][]{Auvergne2009}. This filtered lightcurve was subtracted from the original lightcurve, and all the points deviating more than three standard deviations of the residuals were discarded. Finally, we computed normal points by binning the data on time intervals having approximately the duration of the orbital period of the satellite, obtaining a lightcurve consisting of 752 points that cover 54 days (see Fig.2). Each normal point was acquired by averaging 12 observations with a 512 s sampling.

We used the method described in \citet{Deeg2009} to quantify the straylight contamination from stars located in the vicinity of CoRoT\,102899501. With the use of $BVr'i'$ images collected with the Wide Field Camera at the Isaac Newton Telescope \citep{Deleuil2009}, a reproduction of the CoRoT point spread function was folded over the positions of CoRoT\,102899501 and neighbour stars while accounting for their brightness. We found that the light contamination factor is negligible, being less than 0.1\,\%.

\subsection{Groundbased follow-up spectroscopy}
\label{Sec:Spec FU}

\begin{figure*}[!ht]
\begin{center}
\begin{tabular}{c c}
\psfig{figure=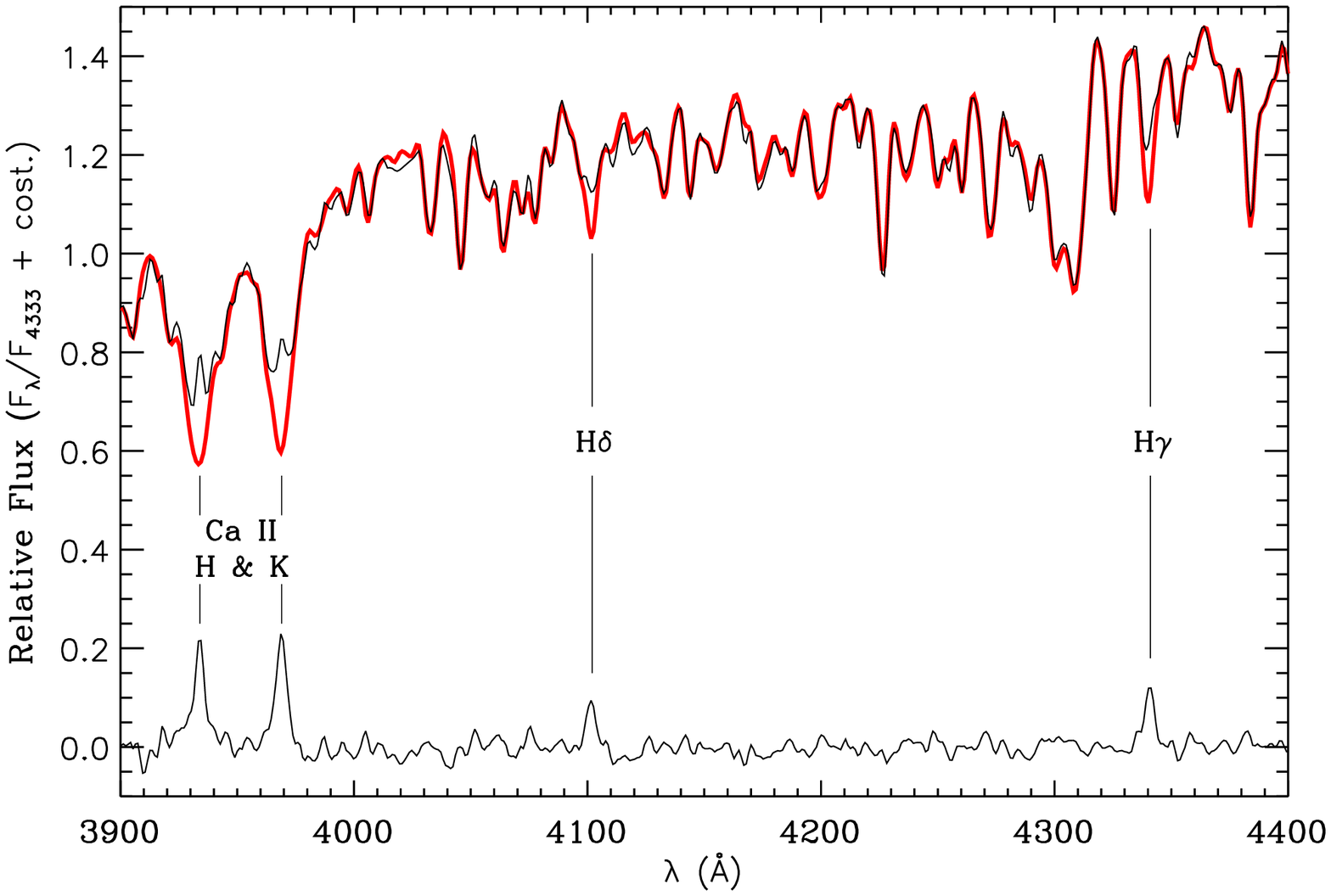,width=8.6cm,angle=0} 
\psfig{figure=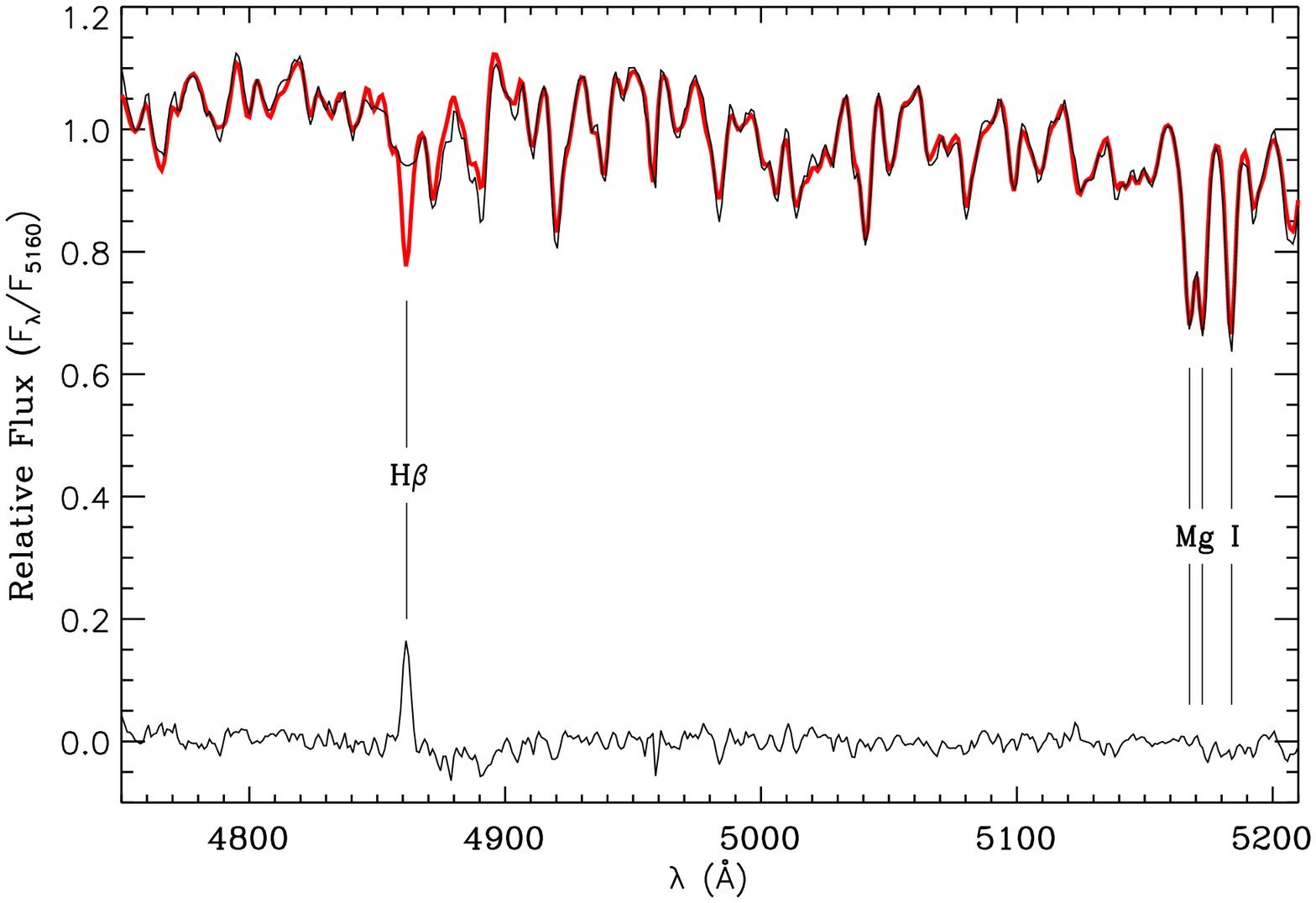,width=8.6cm,angle=0}\\
\psfig{figure=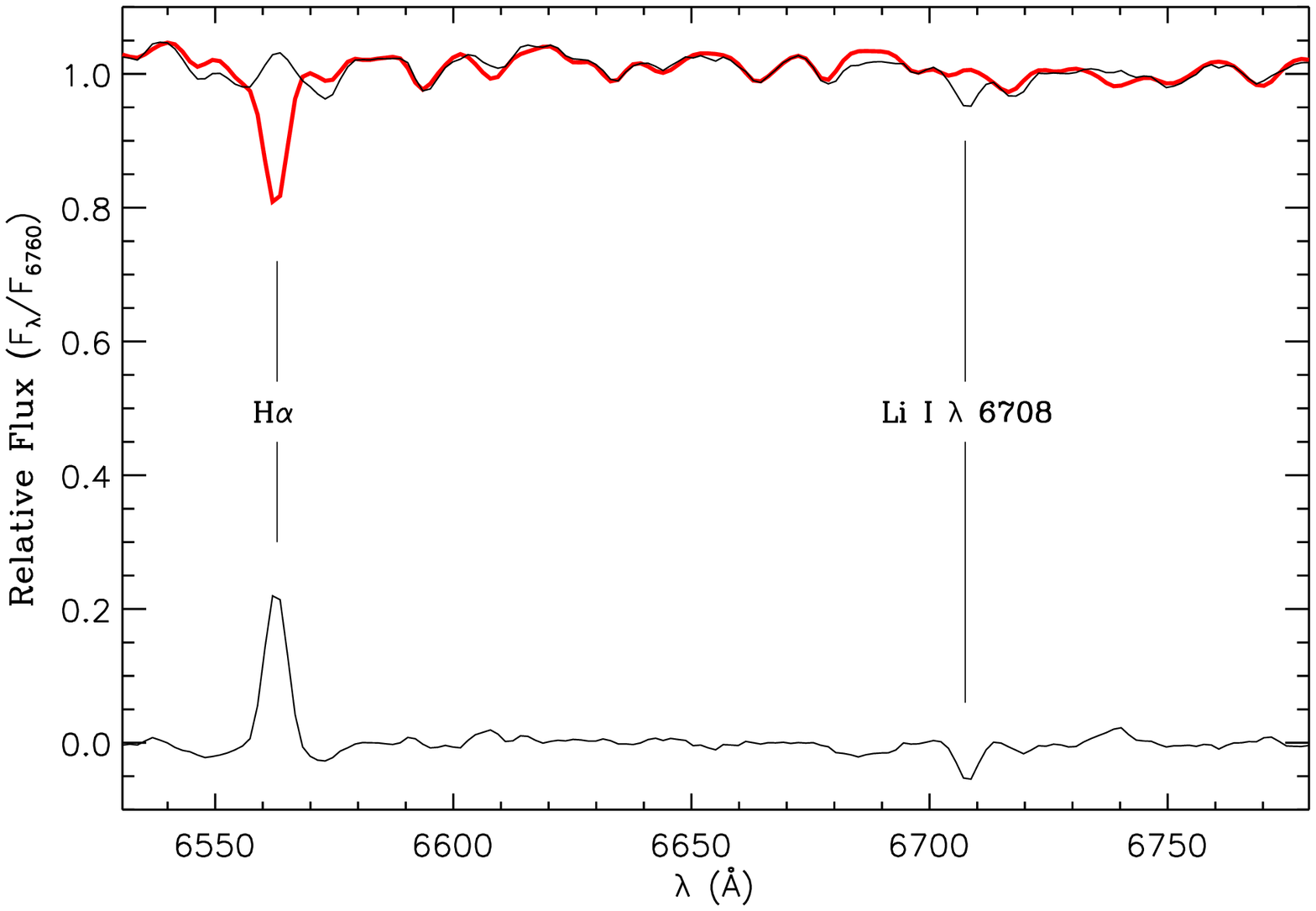,width=8.6cm,angle=0}
\end{tabular}
\caption{AAOmega spectrum of CoRoT\,102899501 (thin line) in three different spectral regions, along with the best-fitting template overplotted (thick line). The spectra have been arbitrarily normalised to the flux reported in the title of the vertical axis. The differences between observed and best-fitting spectrum are displayed in the lower part of each panel. Emission in the Balmer and \ion{Ca}{ii} H\,\&\,K lines is detected, as well as a strong \ion{Li}{i} $\lambda$\,6707.8\,\AA\ absorption line, suggesting a young and active solar-like star.}
\label{Spectrum}
\end{center}
\end{figure*}

\begin{table*}[!t]
  \centering 
  \caption{Radial velocities of CoRoT\,102899501 as obtained with FIES and Sandiford spectrographs.}
  \label{Table_rv}
\begin{tabular}{cccccc}
\hline
\hline
\noalign{\smallskip}
    Date (UT)     &      HJD         &    RV~~ &$\sigma_{RV}$~~&  S/N per pixel & Instrument     \\
  (yyyy/mm/dd)    &     (days)       &  (\kms) &  (\kms)       &  at 5500~\AA)  &                \\
\noalign{\smallskip}
\hline
\noalign{\smallskip}

 2010/10/10       & 2455479.68731698 & 22.013  &     0.252     &      55        & FIES \\
 2010/10/28       & 2455497.65239003 & 21.460  &     0.350     &      13        & FIES \\
 2010/11/01       & 2455501.71001866 & 21.496  &     0.260     &      32        & FIES \\
 2011/01/19       & 2455581.38369968 & 21.516  &     0.297     &      22        & FIES \\
 2011/01/23       & 2455584.83710008 & 21.655  &     0.330     &      26        & Sandiford \\
 2011/01/24       & 2455585.87036894 & 21.598  &     0.356     &      16        & Sandiford \\
 2011/01/28       & 2455589.67732107 & 21.795  &     0.322     &      27        & Sandiford \\
 2011/01/29       & 2455590.72633521 & 21.250  &     0.342     &      28        & Sandiford \\
\noalign{\smallskip}
\hline
\end{tabular}

\end{table*}

A reconnaissance low-resolution ($R\approx1300$) spectrum of CoRoT\,102899501 was acquired with the AAOmega multi-object facility \citep{Sharp2006} at the Anglo-Australian Observatory in January 2009, as a part of the project devoted to study the stellar populations in the \corot\ exoplanet fields \citep{Sebastian2012, Guenther2012}. Spectral type and luminosity class of the target star were derived by comparing the observed AAOmega spectrum with a grid of suitable templates, as described in \citet{Frasca2003} and \citet{Gandolfi2008}. We found that CoRoT\,102899501 is a K0\,V star.

Figure\,1 shows the best-fitting template (thick line) superimposed on the AAOmega spectrum of CoRoT\,102899501 (thin line) for three spectral regions encompassing different lines: the \ion{Ca}{ii} H\,\&\,K, H$\delta$ and H$\gamma$ lines (upper left-hand panel), the H$\beta$ line and \ion{Mg}{i} triplet (upper right-hand panel), and the H$\alpha$ and \ion{Li}{i} 6708\,\AA\ lines (lower panel). A clear emission in both the Balmer and \ion{Ca}{ii} H\,\&\,K lines is detected, confirming the high magnetic activity level suggested by the large amplitude of the \corot\ light curve. A deep \ion{Li}{i} $\lambda$\,6707.8\,\AA\ absorption line, well resolved from the nearby \ion{Ca}{i} $\lambda$\,6718\,\AA\ line, is also visible in the spectrum. By subtracting the best-fitting template from the observed spectrum, we estimated that the equivalent width (EW) of the \ion{Li}{i} $\lambda$\,6707.8\,\AA\ is $\sim300$\,m\AA.

Measurements of the star's radial velocity (RV) were performed to determine whether this rapidly rotating object is a single star or a member of a close binary system with tidally locked components. To this aim, we acquired four high-resolution spectra with the FIES fibre-fed echelle spectrograph \citep{Frandsen1999} attached to the 2.56\,m Nordic Optical Telescope in La Palma (Spain) in October 2010 and January 2011, under the observing programs P40-418 and P42-216. The \emph{Med-Res} fibre was used, yielding a resolving power of $R=47\,800$ in the spectral range 3700-7300\,\AA. We also acquired FIES template spectra of non-active, lithium-poor stars with the same spectral type as CoRoT\,102899501 in December 2011 and January 2012, under observing programs P44-117 and P44-206, to determine the chromospheric activity level and photospheric lithium abundance of the target star (see Sects.~\ref{Sec:Chrom} and \ref{Sec:Lithium}). In January 2011 we gathered four additional high-resolution spectra with the Sandiford cass-echelle spectrometer \citep{Sandiford1993}, mounted at the 2.1\,m (82\,inch) Otto Struve Telescope of McDonald Observatory, Texas (USA). The spectra cover the wavelength range 5000-6000\,\AA\ with a resolving power of $R=47\,000$. Long-exposed ThAr spectra were acquired right before and after each FIES and Sandiford science spectrum to account for RV shifts of the instruments. The data were reduced using IRAF standard routines. We obtained RV measurements by cross-correlating the extracted science data with the spectrum of the radial velocity standard star HD\,50692 \citep{Udry1999}, observed with the same instrument set-up.  

The FIES and Sandiford spectra reveal a single-peaked cross-correlation function with a relatively broad full-width at half maximum FWHM=62\,\kms, corresponding to a projected rotational velocity (\vsini) of about 35\,\kms. This is in agreement with the rapid rotation inferred from the \corot\ lightcurve and excludes a pole-on view of the star. If CoRoT\,102899501 were a tidally locked binary system in a short period orbit (P=1.65 days), its orbital angular momentum vector would be aligned with the stars' rotation spin axis. The system would thus be expected to have a variable RV component along the line of sight with an amplitude of several \kms. Although the accuracy of the RV measurements of CoRoT\,102899501 is affected by the high rotation rate of the star, such a variable RV component is not detected (Table~\ref{Table_rv}). This excludes the presence of a short-period stellar companion to CoRoT\,102899501 with a high confidence level. 

We used the co-added FIES and Sandiford spectra to derive effective temperature ($T_\mathrm{eff}$), surface gravity (log\,$g$), metallicity ($[\rm{M/H}]$), and projected rotational velocity (\vsini) of CoRoT\,102899501. Following the procedure usually adopted in \corot\ exoplanets' discovery papers \citep[e.g.][]{Fridlund2010, Gandolfi2010}, we compared the co-added FIES and Sandiford spectra with a grid of synthetic model spectra from \citet{Castelli2004,Coelho2005}, and \citet{Gustafsson2008}. We also employed the spectral analysis packages SME 2.1 \citep{Valenti1996, Valenti2005}, as well as a modified version of the ROTFIT code \citep{Frasca2003}, which compares observed data with a set of template spectra of real stars with well-known parameters. Consistent results were obtained, regardless of the spectrum and method used. The final adopted values are $T_\mathrm{eff}=5180\pm80$\,K, log\,$g=4.35\pm0.10$\,dex (CGS), $[\rm{M/H}]=0.05\pm0.07$\,dex, and $\vsini=36\pm1$\,\kms (Table 3), in agreement with the spectral type determination obtained from the AAOmega spectrum.

\begin{table}[!t]
\begin{center}
\label{Table_Param}
\caption{Stellar parameters of CoRoT\,102899501 derived from the AAOmega, FIES, and Sandiford spectra.} 
\begin{tabular}{ll}
\hline
\hline
\noalign{\medskip}
Spectroscopic & \\
parameters    & \\
\noalign{\medskip}
\hline
\noalign{\medskip}

$T_\mathrm{eff}$               & $   5180\pm80$\,K     \\
log\,$g$                       & $   4.35\pm0.10$\, \\
$[\rm{Fe/H}]$                  & $   0.05\pm0.07$\, \\
\vsini                         & $   36.0\pm1.0$\,km s$^{-1}$ \\
Sp.T.                          &        K0\,V         \\

\noalign{\medskip}
\hline
\end{tabular}
\bigskip
\end{center}
\end{table}

\begin{figure*}[!ht]
\begin{center}
\begin{tabular}{c}
\psfig{figure=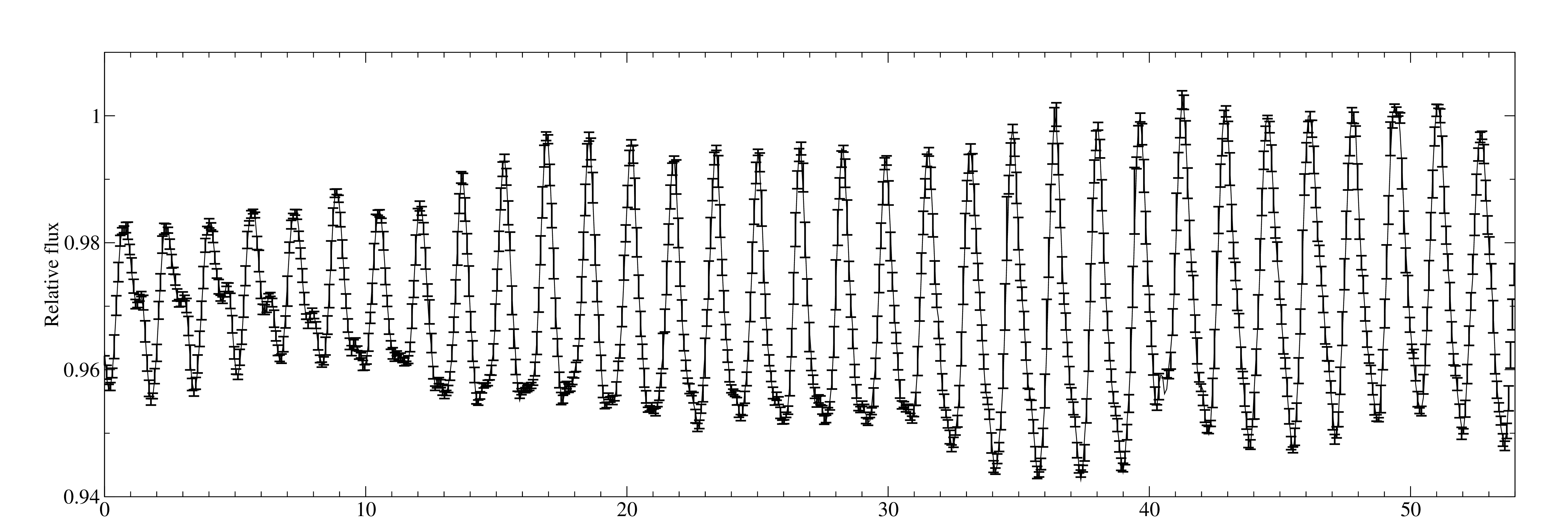,width=6.1cm,angle=270} \\
\psfig{figure=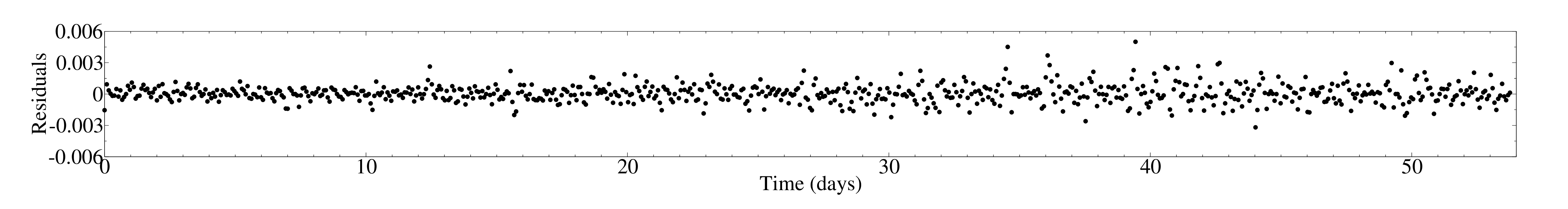,width=2.44cm,angle=270} \\
\end{tabular}
\caption{Best-fit model ($\chi^{2}$=0.90 for 407 degrees of freedom) to the lightcurve of CoRoT\,102899501. The residuals to the best fit are shown in the lower panel.}
\label{Lightcurve}
\end{center}
\end{figure*}

\section{Data analysis}

\subsection{Stellar variability model}
\label{Sec:Stellar variability model}

The rotation modulation of stellar photometric lightcurves by stellar active regions can be modelled using two numerical approaches: surface integration methods and the analytical method \citep{Ribarik2003}. The former assigns a temperature to each pixel of the spherical integration net and then varies each value until an optimal fit to the data is achieved. In the present study, we used the analytical approach described by \citet{Lanza2006}, which is based on a model used to fit the time variations of the solar bolometric and spectral irradiance. Spots and faculae are modelled as point-like sources with flux contributions that account for their area and contrast. Although maximum-entropy regularised spot models show the best agreement with solar observations  \citep{Lanza2007}, discrete spot models constrain spot longitudes and, as a consequence, the differential rotation. Following these models, the variation of the monochromatic stellar flux due to discrete active regions is given by

\begin{equation}
\Delta F(\lambda, t) = \Sigma_{k} \; \mu_{k} \; A_{k}\; I(\lambda,
\mu_{k}) \; \; c_{s}(\lambda),
\end{equation}

\noindent where $\Delta$ $F(\lambda, t)$ is the perturbed stellar flux at wavelength $\lambda$ and time $t$, and $\mu_{k}$ = cos$\psi_{k}$, with $\psi_{k}$  the angle between the normal to the k-th active region and the line of sight. $A_{k}$ is the area of the cool spots in the k-th active region. $I(\lambda, \mu_{k})$ is the specific intensity of the unperturbed photosphere (which depends on $\mu$ owing to limb darkening), and $c_{s}$ is the contrast of the cool spots (see Eq.\,4). Because of our ignorance of the structure of stellar active regions, the facular contribution was neglected based on the analysis results of the lightcurves of active dwarfs \citep{Gondoin2008}. The summation in Eq.\,1 is extended over the active regions on the visible hemisphere, i.e. for which $\mu_{k}$ $>$ 0. The value of $\mu_{k}$ is a function of time that is given by

\begin{equation}
\mu_{k} = \cos i \; \sin \theta_{k} \; + \; \sin i \; \cos \theta_{k}
\; \cos(\Omega_{k} \; t  + \Lambda_{k}),
\end{equation}

\noindent where $i$ is the inclination of the stellar rotation axis along the line of sight, $\theta_{k}$ is the latitude, $\Lambda_{k}$ the longitude, and $\Omega_{k}$ = 2$\pi$/$P_{k}$ the angular velocity of the k-th  active region having a rotation period $P_{k}$.  The specific intensity of the undisturbed photosphere is

\begin{equation}
I(\lambda, \mu_{k}) = {4 \; (\; a_{\rm p}+b_{\rm p} \;\mu + c_{\rm
    p} \; \mu^{2}) \over a_{\rm p}+2b_{\rm p}/3+c_{\rm p}/2} \; B(\lambda, T_{\rm eff}),
\end{equation}

\noindent where the a$_{\rm p}$,  b$_{\rm p}$, and c$_{\rm p}$ are the quadratic limb-darkening coefficients.  Stellar oscillations are not taken into account in the lightcurve simulations that focus on a low-frequency domain of stellar variability. The effects of super-granulation, meso-granulation, and granulation are also neglected. The coefficients of the quadratic limb-darkening law adopted to describe the unperturbed bolometric specific intensity of the stars were derived from \citet{Claret2000}. The contrasts of the cool spots are assumed to be independent of their position on the stellar disk and are estimated as

\begin{equation}
c_{s}(\lambda) = {B(\lambda, T_{\rm s}) \over B(\lambda, T_{\rm eff})} - 1,
\end{equation}

\noindent where $B(\lambda, T)$ is the Planck function, $T_{\rm s}$ is the spot effective temperature, and $T_{\rm eff}$ is the effective temperature of the unperturbed photosphere. The temperatures of starspots estimated using the Doppler imaging technique \citep[e.g.][]{Strassmeier2003} or from the variation of colour indexes vs rotation \citep[e.g.][]{Eaton1992} are between 600 and 1600 K cooler than the unperturbed photosphere for most active stars. \citet{Berdyugina2005} showed that, on average, this temperature difference appears larger for hotter stars, with values near 2000 K for the late F and early G stars dropping to 200 K for the late M stars \citep{Strassmeier2009}. A value $T_{\rm s}-T_{\rm eff} \approx 1500$\, K is found when $T_{\rm eff} = 5300$\, K.  In the model, we assumed $T_{\rm s}= 3800$\, K .

The rotation period found by Fourier analysis is $P_{\rm rot}=1.62\pm0.05$~days in agreement with \citet{Debosscher2009}, with the uncertainty limited by the finite time duration of the lightcurve. Since starspots are used as tracers of the star rotation and can migrate in longitude, this value was refined by modelling the lightcurve itself with a three-spot model (see Sect.\,\ref{Sec:Lightcurve analysis}).   

\begin{table}[!t]
\begin{center}
\caption{CoRoT\,102899501 parameters used in the light curve analysis. The limb-darkening parameters were derived from Claret (2000) using a quadratic limb-darkening law for an LTE stellar model with effective temperature and gravity derived from the spectral analysis.} 

\begin{tabular}{l c c}
\hline
\hline
\noalign{\medskip}
Model Parameters & \\
\noalign{\medskip}
\hline
\noalign{\medskip}
Limb darkening & $a_{\rm p}$ & 0.2005 \\
Limb darkening & $b_{\rm p}$ & 0.9905 \\
Limb darkening & $c_{\rm p}$ & -0.1910 \\
Rotation period & $P_{\rm rot}$ (d) & 1.62 \\
Spot temperature & $T_{\rm s}$ (K) & 3800 \\
Ratio of faculae to spots area & $Q$ & 0 \\
Fitting time & $\Delta$ $t_{\rm f}$ (d) & 1.45 \\ 
\noalign{\medskip}
\hline
\end{tabular}
\bigskip
\end{center}
\end{table}

\begin{figure*}
\begin{center}
\begin{tabular}{c c}
\psfig{figure=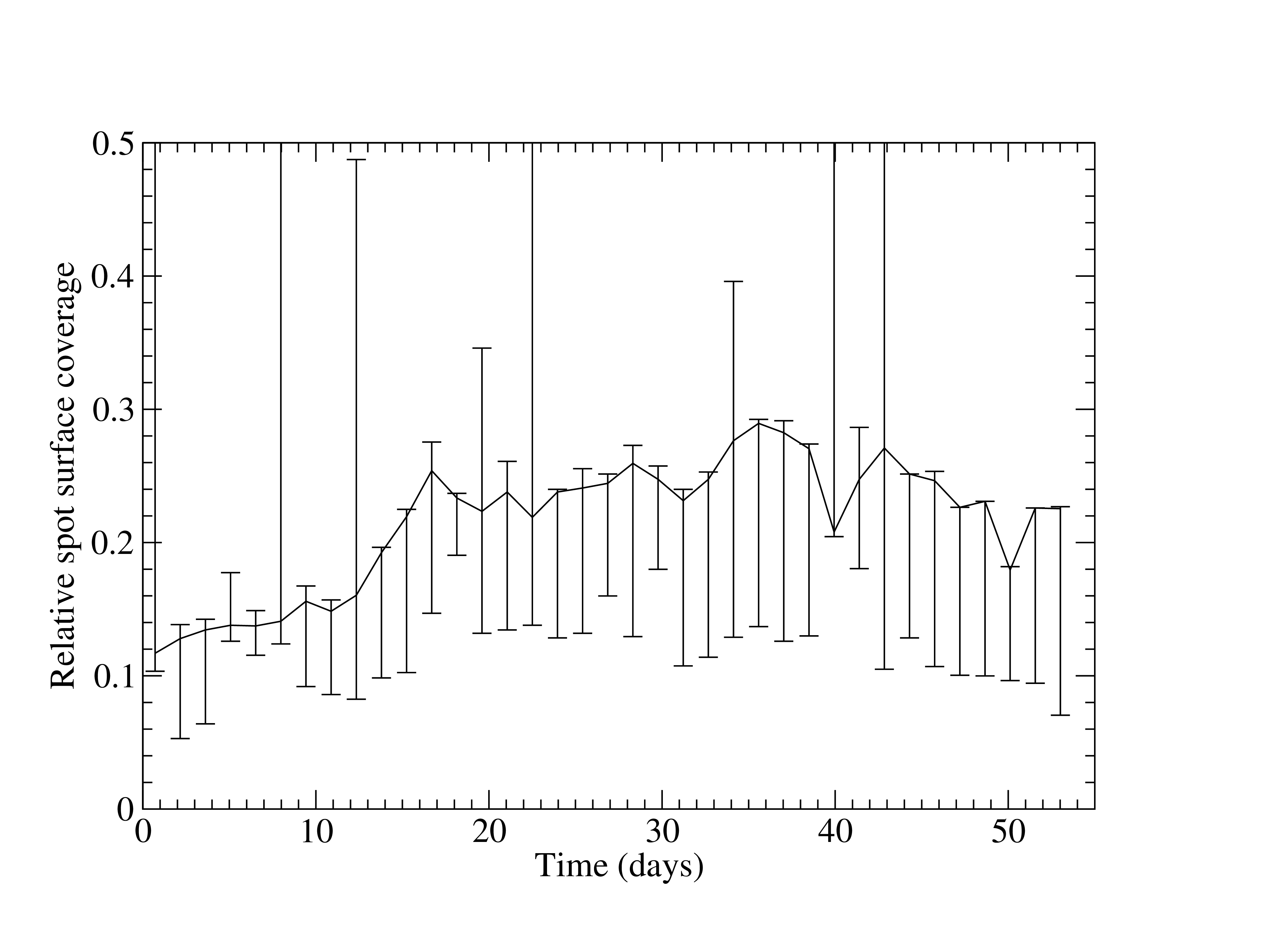,width=7.1cm,angle=270} & \psfig{figure=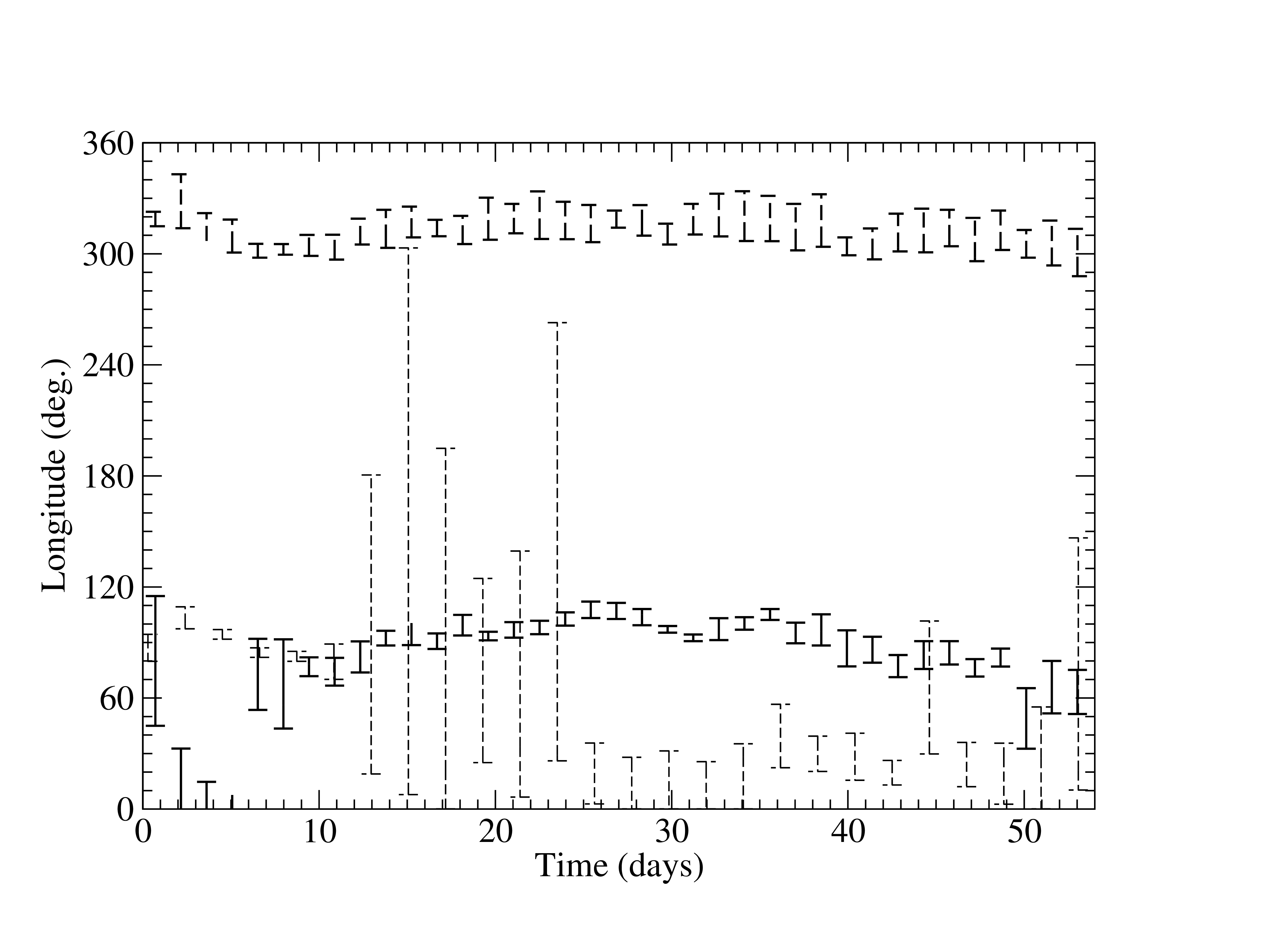,width=7.1cm,angle=270} \\
\end{tabular}
\caption{Time evolution of the total spot surface coverage (left) and active longitude (right) on CoRoT\,102899501 derived from its light curve analysis using a three-spot model. The three sets of symbols on the right-hand side correspond to the three spots used in the model.}
\label{StarSpotCover_DiffRot}
\end{center}
\end{figure*}

\subsection{Lightcurve analysis}
\label{Sec:Lightcurve analysis}

The relative flux variations of the sample star CoRoT\,102899501 were fitted with the three-spot model described above. Best-fit models to the lightcurve were obtained by minimizing the sum of squared residuals. The fixed parameters in the simulations included the stellar and active region parameters previously described. The surfaces, longitudes, and latitudes of the active regions were left as free parameters. The analyses of the lightcurve were performed using a three-spot model and iterating on nine variable parameters. 

It was possible to obtain a good fit of the irradiance changes only for a limited time interval $\Delta$ $t_{\rm f}=1.45$~days, i.e. about 90\,$\%$ of CoRoT\,102599801 rotation period. Hence, the lightcurve was divided in 37 equal intervals covering 54~days of the time series. Each individual sub-lightcurve of 1.45 days' duration was fitted with the three-spot model to derive within each time interval average spot surfaces, latitudes, and longitudes. This method enabled us to estimate the time evolution of each spot parameter along the 54-day duration of the CoRoT\,102599801 lightcurve (see Fig.3). Using the same approach for modelling the rotation modulation of the Sun over several years, \citet{Lanza2003} found that the longest time interval that can be modelled with three stable active regions is 14~days, i.e about 50\,\% of the Sun rotation period. This was the lifetime of the sunspot group dominating the solar irradiance variations. In the case of other active stars, the value of $\Delta$ $t_{\rm f}$ is determined from the observations themselves, looking for the maximum data extension that allows for a good fit. In the case of the active star CoRoT-Exo-2 \citep{Lanza2009}, the maximum time interval $\Delta$ $t_{\rm f}$ that could be fitted with a three-spot model turned out to be 3.2 days, i.e 70$\%$ of the star rotation period. 

Many series of fitting processes varying the inclination of the star rotation axis were conducted in order to identify the inclination angle that minimizes the overall $\chi^{2}$ and the fitting parameters, for which the difference in $\chi^{2}$ becomes significant at more than 99.99$\%$ confidence level. From the fits performed with a range of inclination angles, we kept those that gave the best fits within a 99.99$\%$ confidence level, using them to estimate a range in the other parameters of interest, i.e. spot areas, latitudes, and longitudes. An inclination angle $i = 87^{+1}_{-13}$ degrees was found with this new method. The consistency between the derived inclination angle and the spectroscopically measured projected equatorial velocity is addressed in Sect.4. This range of inclination angles leads to uncertainties in the determination of spots parameters.

By modelling the rotational modulation of the total solar irradiance with the same method, \citet{Lanza2007} noted that the only quantities that can be safely derived are the longitudinal distribution of the active regions and the variation of their total area, measured with respect to a reference value corresponding to a given value of the irradiance assumed to be that of the unperturbed star. The estimated time evolution of the total spot surface coverage and of the active longitude on the star during the 2007 \corot\ initial run are displayed  in Fig.\,\ref{StarSpotCover_DiffRot}. The analysis results do not show a conclusive variation of the spots' surface coverage which, according to the model, is included between $\sim$5-14\,$\%$ at the beginning of the observing run and $\sim$13-29\,$\%$, 35 days later.

For comparison, Doppler images of active stars have shown starspots with a size up to 20\% of a hemisphere \citep{Strassmeier2009}. High filling factors, up to 50\% of the stellar disk, have been determined from modelling molecular bands observed in the spectra of spotted stars \citep{ONeal1996, ONeal1998}. In particular, for the young G1.5 dwarf EK Dra that has been used as a proxy for the young Sun, \citet{ONeal2004} derived a spot temperature of about 3800 K and a filling factor varying between 25\% and 40\%.

Figure\,\ref{StarSpotCover_DiffRot} (right-hand panel) indicates that the active regions on the star photosphere experience no significant drift in longitude. Conversion of the best linear fit to these time evolutions of the longitudes between days 15 and 53 into rotation periods give values of $1.6266\pm0.0008$, $1.6240\pm0.0005$, and $1.6262\pm0.0009$ days for each of the three spots, respectively. The uncertainties on these rotation periods provide no conclusive indication of differential rotation on the surface of the star.

\begin{figure*}[th]
\centering
\includegraphics[width=15cm]{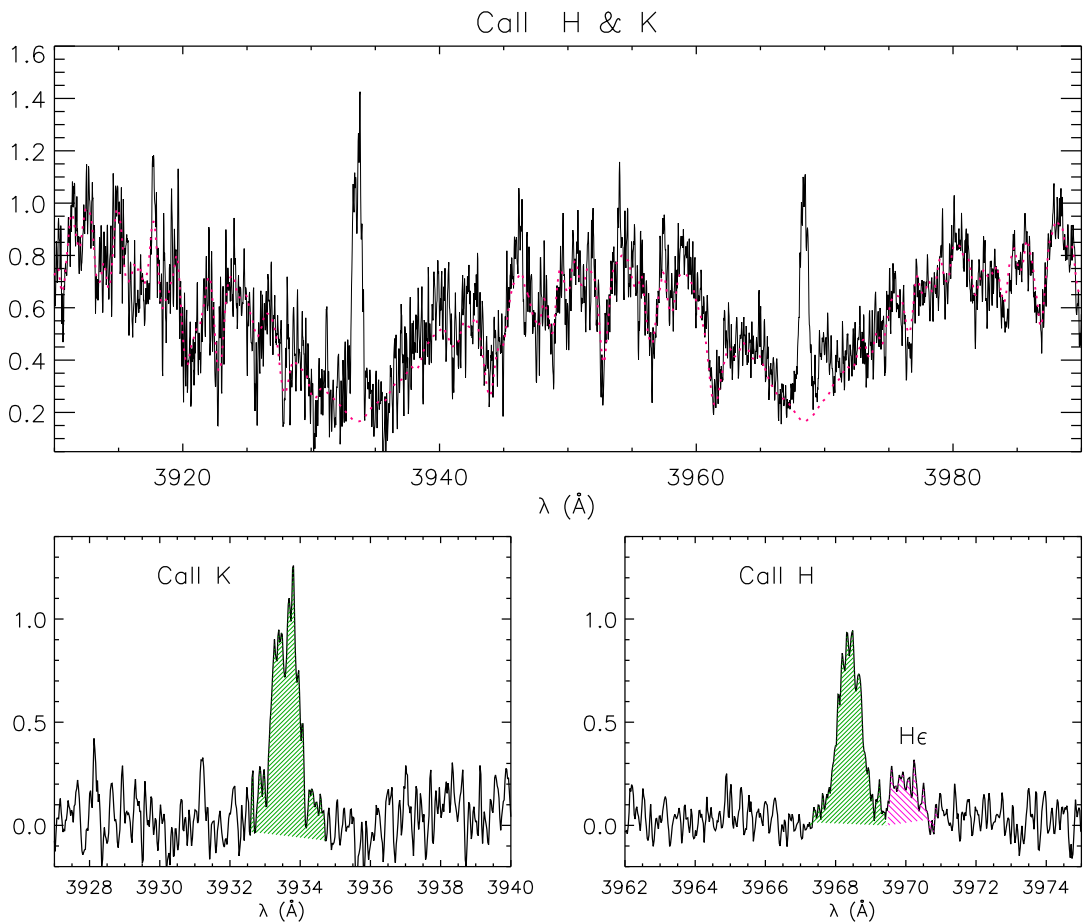}
\caption{{\it Top panel}: continuum-normalized spectrum  of CoRoT\,102899501 (solid line) in the \ion{Ca}{ii} H\,\&\,K region observed on 2010 Oct. 10. The spectral template of the non-active star broadened at the \vsini\ of the target and Doppler-shifted according to the RV difference is overplotted with a dotted line. {\it Bottom panels}: difference spectrum where the H$\epsilon$ emission is emphasized.} 
\label{Fig:CaIIHK}
\end{figure*}

\begin{figure*}[th]
\centering
\includegraphics[width=19.1cm]{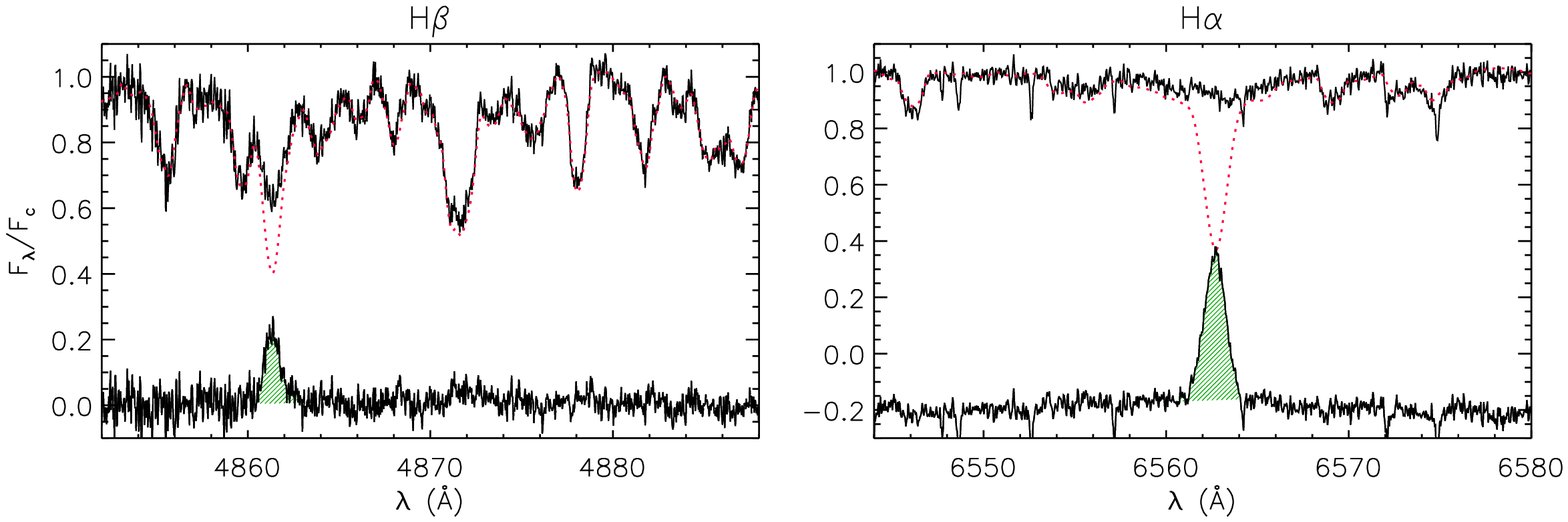}
\includegraphics[width=19.1cm]{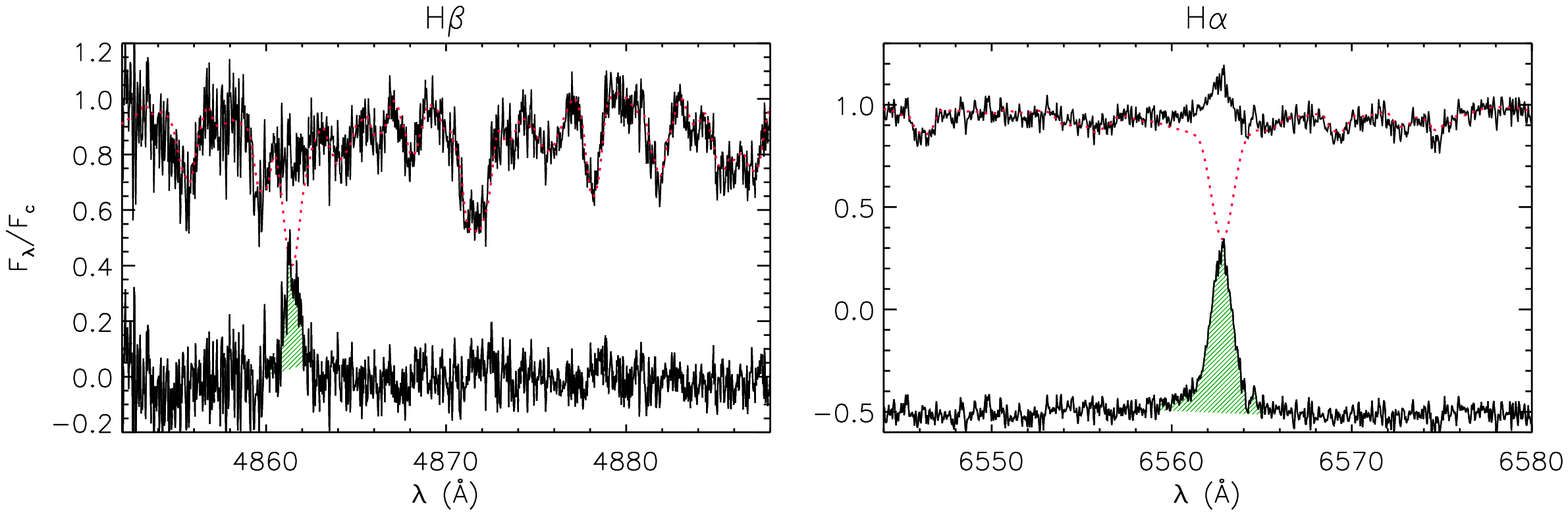}
\caption{{\it Top of each panel}: observed, continuum-normalized FIES spectra of CoRoT\,102899501 (solid line) acquired on 2010 October 10 ({\it upper panels}) and 2010 November 1 ({\it lower panels}) in the H$\beta$ ({\it left panels}) and H$\alpha$ ({\it right panels}) regions. The spectrum template of the non-active star is represented in dotted lines. {\it Bottom of each panel}: difference between observed and template spectra. The residual H$\alpha$ profile is shifted downwards for the sake of clarity. The hatched areas represent the excess emissions that have been integrated to obtain the net line $EW$.} 
\label{Fig:Halpha}
\end{figure*}

\subsection{\ion{Ca}{ii} H\,\&\,K and Balmer lines emission}
\label{Sec:Chrom}

We used the FIES spectra with the highest signal-to-noise ratios acquired on 10 October and 1 November 2010 to measure the emission in the cores of the \ion{Ca}{ii} H\,\&\,K and in the H$\alpha$, H$\beta$, and H$\epsilon$ Balmer lines using the spectral subtraction technique \citep[see, e.g.][]{Herbig1985,Frasca1994}. The method consisted of subtracting a reference spectral template of a non-active star. This template was obtained by a rotational broadening of the observed spectrum of a slowly rotating star with the same spectral type as CoRoT\,102899501, but exhibiting no sign of magnetic activity. The net equivalent widths ($EW$) of the \ion{Ca}{ii} H\,\&\,K, H$\alpha$, H$\beta$, and H$\epsilon$ lines (see Table~\ref{tab:fluxes}) were measured by integrating the residual emission profile in the subtracted spectra (see Figs.~\ref{Fig:CaIIHK} and \ref{Fig:Halpha}).

The \ion{Ca}{ii} H\,\&\,K lines display strong and fairly broad emission cores (Fig.~\ref{Fig:CaIIHK}). A line reversal with a slight asymmetry is visible only for the K line. while the H line does not show such behaviour owing to its lower signal-to-noise ratio. The H$\epsilon$ emission is barely visible in the observed spectrum, but appears clearly after subtraction of the non-active template (Fig.~\ref{Fig:CaIIHK}). The peak intensity of the \ion{Ca}{ii} H\,\&\,K lines is comparable to that observed in \object{KIC\,8429280}, a very young K2 star recently studied by \citet{Frasca2011}.

Fig.~\ref{Fig:Halpha} (top) shows that, on 10 October 2010, the H$\beta$ line is partially filled in with emission, while the H$\alpha$ line is completely filled in with emission (see Table~\ref{tab:fluxes}). The spectrum acquired on 1 November 2010 suggests an even higher activity level, since the H$\beta$ line is filled in and the H$\alpha$ line exhibits an emission profile above the continuum. Similar behavior has been observed on the very young K2 active stars, KIC\,8429280 and \object{LQ\,Hya}, whose H$\alpha$ lines vary from filled in to weak emission profiles \citep[e.g.][]{Strassmeier1993, Frasca2008}.

We evaluated the radiative losses associated with the line excess emission following the guidelines of \citet{Frasca2010}, i.e. by multiplying the average $EW$ by the continuum surface flux at the wavelength of the line. The latter was evaluated by means of the spectrophotometric atlas of \citet{Gunn1983} and the angular diameters calculated by applying the \citet{Barnes1976} relation. The $EW$ and fluxes of the chromospheric lines are reported in Table~\ref{tab:fluxes}. The $EW$ and emission flux of the \ion{Ca}{ii} K line of CoRoT\,102899501 are comparable to those of chromospherically active binaries with similar rotation periods \citep{Montes1996}.  

\begin{table}
\caption{Line equivalent widths and associated radiative losses.}
\centering
 \begin{tabular}{lccc}
  \hline\hline
  \noalign{\smallskip}
  Line                  & Date        & $EW$   & Flux             \\  			     
                        & (yyyy/mm/dd)& (\AA) &(erg\,cm$^{-2}$\,s$^{-1}$) \\
  \noalign{\smallskip}
  \hline
  \noalign{\smallskip}
  H$\alpha$             & 2010/10/10  & 0.872 $\pm$ 0.075 & 4.24$\times10^6$  \\
  H$\alpha$             & 2010/11/01  & 1.376 $\pm$ 0.214 & 6.68$\times10^6$  \\
  \noalign{\smallskip}
  H$\beta$              & 2010/10/10  & 0.256 $\pm$ 0.087 & 1.43$\times10^6$  \\
  H$\beta$              & 2010/11/01  & 0.387 $\pm$ 0.134 & 2.16$\times10^6$  \\
  \noalign{\smallskip}
  H$\epsilon$           & 2010/10/10  & 0.226 $\pm$ 0.150 & 0.58$\times10^6$  \\
  \noalign{\smallskip}
  \ion{Ca}{ii} H        & 2010/10/10  & 0.772 $\pm$ 0.160 & 1.97$\times10^6$  \\
  \ion{Ca}{ii} K        & 2010/10/10  & 0.930 $\pm$ 0.220 & 2.01$\times10^6$  \\
  \noalign{\smallskip}
  \hline
\end{tabular}
\label{tab:fluxes}
\end{table}

On the basis of the H$\alpha$ and H$\beta$ flux, we evaluated a Balmer decrement $F_{\rm H\alpha}/F_{\rm H\beta}\simeq 3.0$ in both FIES spectra. Values of the Balmer decrement in the range 1--2 are typical of optically thick emission by solar and stellar plages \citep[e.g.][]{Buzasi1989,Chester1991}, while prominences seen off-limb give rise to values of  $\sim 10$, which are typical of an optically thin emission source. This suggests that the bulk of the chromospheric emission of CoRoT\,102899501 originates from magnetic regions similar to solar plages and that prominences play a marginal role.

\subsection{Lithium abundance}
\label{Sec:Lithium}

Using the FIES raw spectra, we measured a \ion{Li}{i} $\lambda$\,6707.8\,\AA\ absorption line $EW$ of about 320\,m\AA\, in good agreement with the value inferred from the AAOmega spectrum (Sect.~\ref{Sec:Spec FU}). This value must be corrected for the contribution of the close \ion{Fe}{i} $\lambda$\,6707.4 \AA~line, which is blended with the lithium line of CoRoT\,102899501 due to its large $\vsini$. Adopting the empirical relation proposed by \cite{Soderblom1993}, $\Delta EW_{\rm \ion{Li}{i}} ({\rm m\AA})=20\times(B-V)_0-3$, a corrected value of $EW_{\rm \ion{Li}{i}}\simeq305$\,m\AA\ is found.

\begin{figure}[!t]
\includegraphics[width=8.7cm]{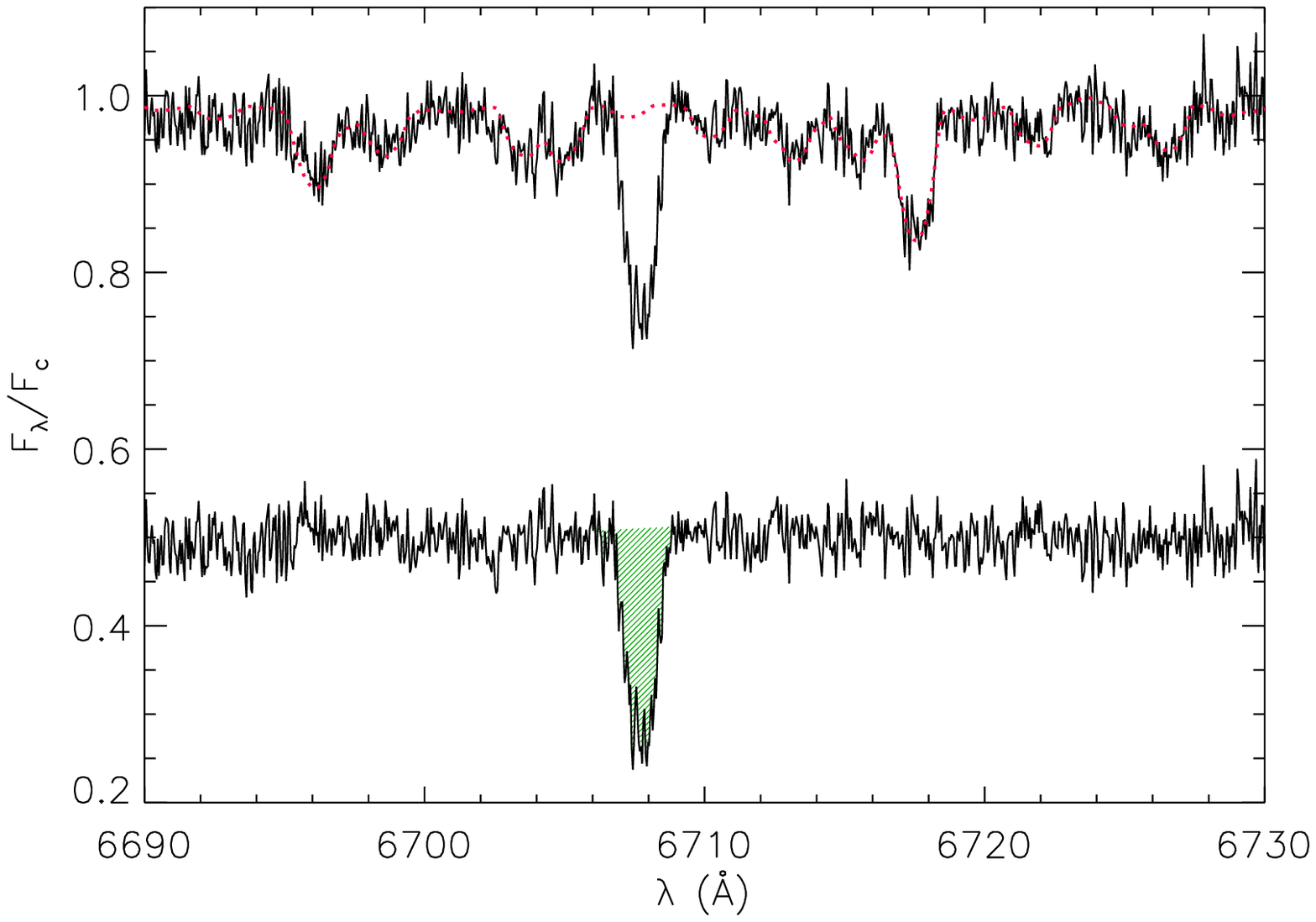}
\caption{Continuum-normalized FIES spectrum of CoRoT\,102899501 obtained on 2010 October 1 in the lithium line region. The template spectrum of a lithium-poor reference star broadened at the $\vsini$ of CoRoT\,102899501 and Doppler-shifted according to the RV difference is shown in dotted line. The lithium absorption is cross-hatched in the difference spectrum whose continuum has been set arbitrarily at a 0.5 level.}
\label{Fig:lithium}
\end{figure}

We also measured the lithium $EW$ on the difference spectrum obtained by subtracting a lithium-poor template broadened at the $v\sin i$ of CoRoT\,102899501 and Doppler-shifted according to the RV difference. As shown in Fig.\,\ref{Fig:lithium}, all the photospheric lines, including the \ion{Fe}{i} $\lambda$\,6707.4\,\AA\  line, are removed in the subtraction, leaving as residuals a lithium absorption with $EW_{\rm \ion{Li}{i}}=295\pm50$\,m\AA .  Adopting the calibrations proposed by \citet{Pavlenko1996}, the lithium line $EW$ translates into a high lithium abundance, $\log N{\rm(Li)}\simeq3.15\pm0.25$. 

\section{Discussion}
\label{Sec:Discussion}

We analysed time-series photometric observations of the star CoRoT\,102899501 observed with the \corot\ space telescope during the initial run IRa01 from 6 February to 2 April 2007. The lightcurve of the star shows an amplitude modulation up to almost 6$\%$ with a period of 1.625~days (see Fig.\,\ref{Lightcurve}). Spectroscopic follow-up observations indicate that CoRoT\,102899501 is a single K0 V star with $T_\mathrm{eff}=5180\pm80$\,K, log\,$g=4.35\pm0.10$, $[\rm{M/H}]=0.05\pm0.07$\,dex, and $\vsini=36\pm1$\,\kms (Table~\ref{Table_Param}). Emissions in both the Balmer line series and the \ion{Ca}{ii} H\,\&\,K lines, as well as the presence of a strong \ion{Li}{i} 6708\,\AA\ absorption line (EW$\approx$295$\pm$50\,m\AA\ ) in the spectrum, suggest that the object is a young single star with a high level of magnetic activity.  The bulk of the chromospheric emission of CoRoT\,102899501 could be due to magnetic active regions similar to solar plages associated with sunspots.

Chromospheric activity has been traditionally measured using the $R_{\rm HK}^{'}$ index, defined as the ratio of the emission in the core of \ion{Ca}{ii} H\,\&\,K lines to the total bolometric emission of the star \citep{Noyes1984}. We derived a value  $log (R_{\rm HK}^{'})$=-4.01$_{-0.14}^{+0.11}$ for CoRoT\,102899501, injecting the measured emission fluxes $F_{\rm H}^{'}$ and $F_{\rm K}^{'}$ in the cores of the \ion{Ca}{ii} H\,\&\,K lines (see Table~\ref{tab:fluxes}) into the following expression \citep{MartinezArnaiz2010}:

\begin{equation}
R_{\rm HK}^{'} = {F_{\rm H}^{'} + F_{\rm K}^{'} \over \sigma T_{\rm eff}^{4}},
\end{equation}

where $\sigma$ is the Stefan-Boltzmann constant. This value is similar to that of the most active solar-type dwarfs among a sample of main sequence and pre-main sequence stars compiled by \citet{Mamajek2008}. These authors show that a chromospheric activity index $log (R_{\rm HK}^{'}) \approx$ -4 is found among solar-type dwarfs that are members of young stellar associations such as Upper Sco (age $\sim$ 5 Myrs), $\beta$ Pic ($\sim$ 12 Myrs), Upper Cen-Lup ($\sim$ 16 Myrs) or Lower Cen Cru ($\sim$ 16 Myrs). Sun-like stars that are members of older open clusters such as $\alpha$ Per ($\sim$ 85 Myrs), the Pleiades ($\sim$ 130 Myrs), UMa ($\sim$ 500 Myrs), the Hyades ($\sim$ 625 Myrs), or M67 ($\sim$ 4000 Myrs) have significantly lower \ion{Ca}{ii} emissions (see Fig.7).

We fitted the relative flux variations of the star's lightcurve with a three-spot model. Assuming a 3800 K spot temperature, the analysis result suggests a large coverage of CoRoT\,102899501 photosphere by active regions. It does not show a conclusive variation of the spots' surface coverage, which, according to the model, is included between $\sim$5-14\,$\%$ at the beginning of the observing run and $\sim$13-29\,$\%$, 35 days later. The starspots used as tracers of the star rotation constrain the rotation period to $1.625\pm0.002$~days and do not show evidence for differential rotation. CoRoT\,102899501 is characterized by a high level of magnetic activity most likely linked to its fast rotation and spectral type.

The effective temperature, gravity, and metallicity derived from the spectroscopic observation of the target (Sect.~\ref{Sec:Spec FU}) were compared with evolutionary models of stars with the same metal abundance. These models were computed by \citet{Siess2000} using the Grenoble stellar evolution code \citep{Forestini1994} and by \citet{Marques2008} using the CESAM code \citep{Morel1997,Morel2008} and the initial condition of the birth line from \citet{Palla1991,Palla1992}. The comparison shows that CoRoT\,102899501 is located near the evolutionary tracks of a 1.09$\pm$0.12~M$_\odot$ pre-main sequence star at an age of 23$\pm$10~Myrs.

The comparison also indicates that the radius of the star is included between 0.96 $R_{\odot}$ and 1.36 $R_{\odot}$. Taking into account the rotation period ($P_{\rm rot} = 1.625\pm0.002$\,days) and the inclination angle ($ 74\degr\ < i < 88\degr\ $) of the star, as derived from the analysis of its lightcurve, we found $\vsini=35.6\pm6.9$\,\kms. The good agreement with the projected equatorial velocity $\vsini=36\pm1$ km s$^{-1}$ inferred independently from the broadening of the spectral lines supports the consistency of the overall analysis.

A 1.625-day rotation period divided by a convective turnover time $\tau_{\rm c}$ = 12.7 days for a 1.09 M$_\odot$ main sequence star \citep{Wright2011} leads to a Rossby number $R_{\rm 0}$ = 0.13. Applying the empirical relation between the $R_{\rm HK}^{'}$ index and the Rossby number (see Eq.(7)) established by \citet{Mamajek2008}, one finds $R_{\rm HK}^{'} = -4.08 \pm 0.03$ in good agreement with the measured emission fluxes in the core of the \ion{Ca}{ii} lines of CoRoT\,102899501.

We followed the method described in \citet{Gandolfi2008} to derive the interstellar extinction towards CoRoT\,102899501. Adopting a normal value for the ratio of total-to-selective extinction ($R_{\mathrm V}=A_{\mathrm V}/E_{\mathrm {B-V}}=3.1)$, we found $A_{\mathrm V}=0.35\pm0.15$~mag and an absorption-corrected star V magnitude equal to 12.49 $\pm$ 0.21. A comparison with the absolute magnitude M$_{\mathrm V}=5.13\pm0.40$ inferred from the evolutionary models leads to a true distance modulus of 7.36$\pm$ 0.61 that corresponds to a distance of 308$\pm$85 pc.

Solar-type stars reach the zero-age main sequence (ZAMS) rotating at a variety of rates, as seen in the Pleiades \citep{Stauffer1987,Soderblom1993}. However, it has been noted \citep[see, e.g.,][]{Barnes2003,Meibom2009} that these young stars tend to group into two main sub-populations that lie on narrow sequences in diagrams where the measured rotation periods of the members of a stellar cluster are plotted against their $B-V$ colours. One sequence, called the $I$ sequence, consists of stars that form a diagonal band of increasing rotation period with increasing $B-V$ colour. In young clusters, another sequence of ultra-fast rotating stars called the $C$-sequence, is also observed, bifurcating away from the $I$-sequence towards shorter rotation periods.

A parameterization of the stellar rotation period as a function of colour index and time is proposed by \citet{Barnes2003} for the two sequences, thus opening the possibility of using stellar gyrochronology for ``gyro-age'' determination. Applying these relationships to CoRot\,102899501 (P$=1.625$\,days; $(B-V)_{\rm 0}=0.77 \pm 0.15$\,mag), we found ages in the range 70-180~Myrs and 8-25~Myrs, using the $C$- and the $I$-sequence relationship, respectively. The age inferred from the $I$-sequence relationship is in good agreement with the 23$\pm$10~Myrs value derived from the stellar evolution models. According to the physical explanation of the $I$-sequence proposed by Barnes (2003), this suggests that the magnetic fields on CoRoT\,102899501, which cause angular momentum loss by coupling the surface of the star to the magnetized wind, are also able to couple to a substantial fraction of the whole star, which could be essentially in solid body rotation.

The C/I dichotomy for stellar rotation was recently formulated mathematically by \citet{Barnes2010} in a simple model that describes the rotational evolution of cool stars on the main sequence. According to this model, the time evolution of the rotational period of a main sequence star depends on two parameters, (i) its initial period of rotation $P_{\rm 0}$ on the ZAMS and (ii) its convective turnover time. These parameters are related by the following expression \citep{Barnes2010}:

\begin{equation}
t = {\tau_{\rm c, B} \over k_{\rm C}} \times \ln({P(t) \over P_{\rm 0}}) + {k_{\rm I} \over 2 \tau_{\rm c, B}} \times (P(t)^{2} - P_{\rm 0}^{2}),
\end{equation} 

where the constants $k_{\rm c}$ = 0.646 days Myrs$^{-1}$, $k_{\rm I}$ = 452 Myr day$^{-1}$, and $\tau_{\rm c, B}$ = 34.884 days have been calibrated on the Sun with input from open-cluster rotation observations, demanding that the rotation of the star starts off with an initial period of 1.1 days and be 26.09 days at the age of 4570 Myrs \citep{Barnes2010}. We combined the above expression with the unique mass-independent prediction of the chromospheric activity index $R_{\rm HK}^{'}$ \citep{Mamajek2008}, expressed as follows:

\begin{eqnarray}
\log R_{\rm HK}^{'} = A - B \times ({P(t) \over \tau_{\rm c}}  - C) \nonumber \\
A, B, C = \left\{ \begin{array}{rl}
  -4.23, 1.451, 0.233 &\mbox{if P(t)/$\tau_{\rm c}$  $<$ 0.32} \\
  -4.522, 0.337, 0.814 &\mbox{if P(t)/$\tau_{\rm c}$ $\ge$ 0.32}.
       \end{array} \right.
\end{eqnarray}

The combination of Eq.(6) and Eq.(7) constitutes a simple time evolution model of the chromospheric activity of main sequence stars as a function of the stars convective turnover time and initial period of rotation $P_{\rm 0}$ on the ZAMS. The dependence on mass of the $R_{\rm HK}^{'}$ index is defined by the parameterization of the convective turnover time as a function of stellar mass provided by \citet{Wright2011}:

\begin{equation}
\log(\tau_{\rm c, W}) = 1.16-1.49 \times \log({M \over M_{\rm \odot}})-0.54 \times \log^2({M \over M_{\rm \odot}}).
\end{equation}

\begin{figure}[!t]
\psfig{figure=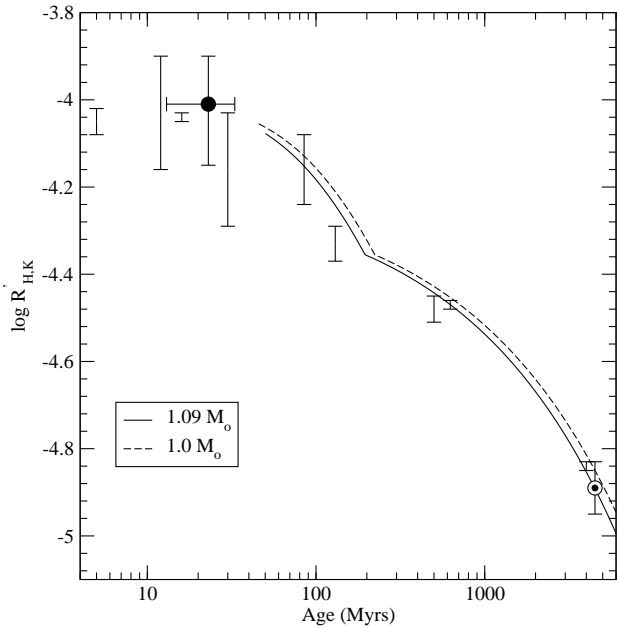,width=9.4cm,angle=0}
\caption{Time evolution model of the chromospheric activity of Sun-like stars on the main sequence compared with the median chromospheric activity indices of the Sun and solar-type dwarfs in Upper Sco (age $\sim$ 5 Myrs), $\beta$ Pic ($\sim$ 12 Myrs), Upper Cen-Lup ($\sim$ 16 Myrs), Lower Cen Cru ($\sim$ 16 Myrs), $\alpha$ Per ($\sim$ 85 Myrs), the Pleiades ($\sim$ 130 Myrs), UMa ($\sim$ 500 Myrs), the Hyades ($\sim$ 625 Myrs), and M67 ($\sim$ 4000 Myrs) compiled by \citet{Mamajek2008}. The value of the chromospheric activity index $log (R_{\rm HK}^{'})$of CoRoT\,102899501 is indicated with a black circle.}
\label{ngc1039:fig:LxvsPM}
\end{figure}

In Eq.(6), we rescaled the turnover time mass dependence of \citet{Wright2011} by a factor ($\tau_{\rm c, B}  / \tau_{\rm c, W}$)$_{\rm \odot}$ $\approx$ 2.4 to correct for the different value of the Sun convective turnover time used by \citet{Barnes2010}. Using the above model, we calculated the time evolution of the $R_{\rm HK}^{'}$ index time of a Sun-like star with an initial period of rotation on the ZAMS and a mass identical to that of  CoRoT\,102899501. The results are compared with the median $log (R_{\rm HK}^{'})$ values of the Sun and of solar-type dwarfs in open stellar clusters with different ages compiled by \citet{Mamajek2008}. The comparison shows that the high level of chromospheric activity of the CoRoT\,102899501 is linked with its fast rotation due to its young age.

In view of the low value of its Rossby number, CoRoT\,102899501 is expected to emit X-rays in the saturation regime of coronal emission at an X-ray to bolometric luminosity ratio $L_{\rm X}/L_{\rm bol} \approx$ 10$^{-3}$ \citep{Pizzolato2003} corresponding to $L_{\rm X} \approx 3 \times 10^{30}$ erg s$^{-1}$. It has been argued that the saturated regime of X-ray emission is associated with a turbulent dynamo \citep{Wright2011} and that a dynamo regime transition, possibly between a turbulent dynamo and an interface-type dynamo, occurs at a Rossby number of 0.3 \citep{Gondoin2012}. This could also explain the change in the  $log (R_{\rm HK}^{'})$ vs rotation relationship observed by \citet{Mamajek2008} around this Rossby number value. CoRoT\,102899501 and the Sun could thus be generating magnetic fields in different dynamo regimes.

A deep \ion{Li}{i} $\lambda$\,6707.8\,\AA\ photospheric absorption line along with magnetic activity indicators as observed in the spectrum of CoRoT\,102899501 are generally considered as a sign of stellar youth \citep[e.g.][]{Soderblom1998}. Lithium is indeed expected to be strongly depleted from the stellar atmospheres of late-type stars when mixing mechanisms pull it deeply in their convective layers. It has thus been suggested \citep[e.g.][]{Skumanich1972} that the lithium surface abundance should decrease with stellar evolution. Recently \citet{DaSilva2009} derived the distribution of lithium abundances in nine stellar associations covering ages from 5\,Myrs up to that of the Pleiades (100\,Myr). These authors measured a systematic decrease of lithium abundance with age in the temperature range from about $3500$\,K to $5000$\,K. They concluded that the age sequence of the young associations based on the lithium abundance measurements agrees well with isochronal age determination. However, they also noticed a scatter of lithium abundance value that hampers the determination of reliable age on individual stars. Also, lithium depletion has been found in several T\,Tauri stars at levels inconsistent with their young ages \citep{Magazzu1992,Martin1994}. 

Although lithium cannot be used as a reliable age indicator for a single star, it is worth noting that the measured $EW$ of the lithium line ($EW_{\rm \ion{Li}{i}}$ = 0.295 $\pm$ 0.050 \AA\ ) in CoRoT\,102899501 spectrum is larger than the upper envelopes of the lithium equivalent widths in the dwarf spectra of both the Pleiades \citep[100\,Myr;][]{Soderblom1993} and $\alpha$\,Per cluster \citep[50\,Myr;][]{Sestito05}. The CoRoT\,102899501 $EW_{\rm \ion{Li}{i}}$ value is actually comparable to that of stars with similar effective temperatures in the young cluster IC\,2602 (30\,Myr, \citealt{Montes2001}). This suggests an age $\le$ 30 Myr consistent with the age determinations inferred from the gyrochronology method and from the comparison with stellar evolution models.

Our analysis of a large set of photometric and spectroscopic observations of  CoRot\,102899501 provides a consistent set of stellar parameters, activity level, and age for a single field star. In particular, the stellar evolution models, lithium abundance, and the gyrochronology technique concur with an age estimate in the range of 8 - 30 Myrs. Moreover, the chromospheric activity of CoRot\,102899501 is consistent with the $R_{\rm HK}^{'}$ index measured on a sample of solar-like dwarfs in young open clusters. These chromospheric activity levels, together with those observed on solar-type dwarfs in older open clusters and on the Sun, are also in line with a model of chromospheric activity evolution on the main sequence that we established by combining  $R_{\rm HK}^{'}$ vs Rossby number relationships with a recent model of stellar rotation evolution on the main sequence \citep{Barnes2010}. We thus conclude that a magnetic activity level comparable to that observed on CoRot\,102899501 could have been present on the Sun at the time of planets' formation.

\begin{acknowledgements}

We are grateful to Dr. A. Baglin and to the \corot\ team for the outstanding quality of the \corot\ data. 
AH, DG, and EG acknowledge the support of the grants 50OW0204 from the Deutschen Zentrums f\"ur Luft- und Raumfahrt (DLR).
The team at the IAC acknowledges support by grant AYA2010-20982-C02-02 of the Spanish Ministry of Economy and Competitiveness.
The authors are grateful to the staff at AAO, NOT, and McDonald Observatories for their valuable support and contribution to the success of the AAOmega, FIES, and Sandiford observing runs.
We are grateful to the anonymous referee for the helpful comments that allowed us to improve the paper.
\end{acknowledgements}


\end{document}